# Hardening and Strain Localisation in Helium-Ion-Implanted Tungsten

Suchandrima Das[a,1], Hongbing Yu[a], Edmund Tarleton[a,b,2], Felix Hofmann[a,3]

[a]*Department of Engineering Science, University of Oxford, Parks Road, Oxford OX1 3PJ, UK*
[b]*Department of Materials, University of Oxford, Parks Road, Oxford OX1 3PH, UK*

[1]*suchandrima.das@eng.ox.ac.uk*

[2]*edmund.tarleton@eng.ox.ac.uk*

[3]*felix.hofmann@eng.ox.ac.uk*

**ABSTRACT**

Tungsten is the main candidate material for plasma-facing armour components in future fusion reactors. In-service, fusion neutron irradiation creates lattice defects through collision cascades. Helium, injected from plasma, aggravates damage by increasing defect retention. Both can be mimicked using helium-ion-implantation. In a recent study on 3000 appm helium-implanted tungsten (W-3000He), we hypothesized helium-induced irradiation hardening, followed by softening during deformation. The hypothesis was founded on observations of large increase in hardness, substantial pile-up and slip-step formation around nano-indents and Laue diffraction measurements of localised deformation underlying indents. Here we test this hypothesis by implementing it in a crystal plasticity finite element (CPFE) formulation, simulating nano-indentation in W-3000He at 300 K. The model considers thermally-activated dislocation glide through helium-defect obstacles, whose barrier strength is derived as a function of defect concentration and morphology. Only one fitting parameter is used for the simulated helium-implanted tungsten; defect removal rate. The simulation captures the localised large pile-up remarkably well and predicts confined fields of lattice distortions and geometrically necessary dislocation underlying indents which agree quantitatively with previous Laue measurements. Strain localisation is further confirmed through high resolution electron backscatter diffraction and transmission electron microscopy measurements on cross-section lift-outs from centre of nano-indents in W-3000He.

## 1. INTRODUCTION

There is an ongoing search for plasma facing armour materials for critical components in future fusion reactors, such as the main wall and the divertor. Tungsten is a promising candidate due to its high melting point, reasonable strength at high temperatures and low sputtering yield [1–3]. Within the fusion reactor, armour components will be exposed to high temperatures (> 1200 K) and irradiated by high energy neutrons (14.2 MeV, ~ 2 dpa per year for tungsten) which will generate lattice defects [3,4]. Foreign elements such as hydrogen isotopes and helium will accumulate within initially pure tungsten components due to transmutation and migration from the fusion plasma [4]. The presence of helium in tungsten is particularly worrying because of its low solubility and strong affinity for lattice defects [5,6].

Ideally, tungsten irradiated in the fusion reactor should be studied to examine the role played by helium in deciding the evolution of its properties. However, such samples, being radioactive, are difficult to access and unsuitable for regular laboratory experiments. Helium-ion implantation can mimic nuclear fusion induced radiation damage by creating irradiation-damage-like lattice defects, while allowing examination of the interaction of the injected helium with these lattice defects. As such helium-implanted tungsten has been widely studied through both simulations and experiments for varying conditions of temperature and helium fluence [7–10]. Studying helium-implanted samples exposed at room temperature allows isolated investigation of the effect of helium-implantation-induced defects on material properties, while excluding the effects of elevated temperature, impurity content, crystallographic orientation etc.

Helium-implantation defects induced at room temperature are challenging to probe. For example, high-resolution Transmission Electron Microscopy (TEM) micrographs of 0.3 at.% helium-implanted tungsten samples (exposed at room temperature) showed no visible damage [11]. This implies that the induced defects are smaller than 1.5 nm, the sensitivity limit of TEM to lattice defects [12]. An interesting question that arises is whether these tiny defect complexes, "invisible" by TEM, can significantly change the physical or mechanical properties of pure tungsten.

Nano-indentation has been widely used to gain mechanistic insight into the shallow helium-damaged layers. Nano-indentation of 0.3 at.% helium-implanted samples (implanted at 298 K) showed a 30% increase in hardness [13,14]. Corresponding observations of 50% lower thermal diffusivity (at room temperature) [15] were also made. Helium-induced changes were also distinctly visible in the indent morphology. Large pile-up and slip steps were noticed around indents in the helium-implanted tungsten sample as compared to shallow pile-up around indents in pure tungsten. This was seen for both Berkovich indents (250 nm deep [13]) and spherical indents (~ 500 nm deep [14]). Moreover, 3D resolved micro-beam Laue diffraction measurements of residual lattice distortions beneath the indents showed a significantly smaller deformation zone in the implanted material [14].

These observations from past studies make it clear that despite being small, the defects induced by helium-implantation at room temperature can significantly alter the physical and mechanical properties of tungsten. However, the mechanisms underlying these observations remain unclear. To prevent these undesirable helium-implantation induced changes, it is essential to understand how the induced defects interact with the surrounding microstructure and evolve. This is also vital for enabling accurate predictions of the evolution of structural and functional properties and the lifetime of irradiated tungsten armour components; a critical aspect for the successful design and operation of commercially viable fusion reactors.

To address the problem, we first need to understand the nature of defects induced by helium-implantation at room temperature. We conducted a prior study on 3000 appm helium-implanted tungsten samples, exposed at room temperature [16]. The nature of implantation defects was deduced in that study using a combination of DFT calculations, spatial scale independent elasticity equations and high resolution mono-beam Laue diffraction measurements (of the order of $10^{-4}$) of the implantation induced residual strain [16]. At room temperature, an energetic helium ion, creates numerous Frenkel pairs. While most Frenkel pairs recombine immediately, a few are prevented from recombining, as helium occupies vacancies to form stable helium-vacancy complexes [6,17]. It was found that the defect microstructure in such cases was dominated by di-helium vacancy complexes, to which a self-interstitial (SIA) remains tightly bound ($He_2V$-SIA complex) [16–18].

We hypothesize that initially $He_2V$-SIA defect complexes can strongly obstruct dislocation motion owing to the strong binding energy of the di-helium-vacancy complex of ~3.1 eV (which makes it difficult to cut through the $He_2V$-SIA obstacle) [6]. This can explain the increase in hardness observed in these samples [13,14]. However, as deformation progresses, we propose that there comes a stage when gliding dislocations can exert sufficient force on $He_2V$-SIA pinning obstacles to dissociate the helium from the helium-vacancy complex. The SIA component of the defect complex may then quickly jump into the vacancy, annihilating the Frenkel-pair which was retained by the helium. This defect removal would facilitate easier glide of subsequent dislocations through defect-free channels and lead to the slip step formation seen near nano-indents in helium-implanted tungsten, as mobile dislocations exit at the free surface [14].

Strain localisation through formation of such slip channels during inhomogeneous plastic deformation has been observed in several past studies on irradiated metals, such as copper [19], steel [20], vanadium [21], zircaloy-4 [22] etc. Indeed, the prediction of the slip channel formation was made by Cottrell [23] even before its first observation in TEM. The earliest experimental observations of dislocation channelling were made by Makin and Sharp, in neutron irradiated single crystals of copper and copper alloys [19,24], based on which they also proposed a mathematical model for slip line and pile-up formation in irradiated metals [19]. Their model is based on a hypothesis similar to ours, whereby irradiation-induced defect clusters are destroyed by gliding dislocations. Similar ideas considering removal of irradiation defects have also been considered in several other studies [25–28]. The objective of this study is to explore this hypothesis for the case of helium-implanted tungsten, using a combination of modelling and experimental techniques. A model built on this hypothesis will capture the physics of the internal damage mechanism. In the long term, such models could be useful for predicting the lifetime of engineering components, or to conduct virtual experiments.

Here we consider a <001>-oriented tungsten single crystal, partly implanted with 3000 appm of helium at room temperature. Nano-indentation with a spherical indenter tip was done on both helium-implanted and unimplanted parts of the sample. The physical morphology of the indent surface and region beneath the indents was examined using atomic force microscopy (AFM), scanning electron microscopy (SEM) and transmission electron-microscopy (TEM) to assert the prior observations of helium-induced changes [14].

To test the hypothesis, a crystal plasticity finite element (CPFE) formulation was developed to simulate indentation in helium-implanted tungsten. A physically-based constitutive law is implemented in the model to effectively capture the hypothesised mechanism of interaction between glide dislocations and helium-implantation induced defects. CPFE predictions of lattice distortions beneath indents are compared to corresponding observations from X-ray micro-beam Laue diffraction and high-resolution electron backscatter diffraction (HR-EBSD) measurements. The spatial distribution of geometrically necessary dislocations (GNDs) beneath indents is calculated from both simulations and experiments as an indicator of the extent of plastic deformation. The results are discussed in light of the proposed mechanism for the interaction of helium-implantation-induced defects and glide dislocations in tungsten.

## 2. RESULTS & DISCUSSION

### 2.1. Load-displacement curves, indent morphology and defect structures

Figure 1 (a)-(b) show SEM micrographs of spherical nano-indents in the unimplanted and implanted sample respectively. To ensure reproducibility, two indents are made in the helium-implanted sample as shown in Figure 1 (b). In striking contrast to the unimplanted sample, a large, localised pile up is seen around both indents in the implanted material. Black arrows in Figure 1 (b) indicate slip steps, sugestive of slip localisation and slip channel formation. Figure 1 (c) shows the load versus displacement curves for nano-indentation to 500 nm depth in helium-implanted and unimplanted parts of the <001>-oriented tungsten single crystal. The implanted sample is seen to reach ~ 73% higher load than the unimplanted sample. This substantial increase in hardening may be attributed to both the helium-implantation induced defects, as well as the increase in pile-up, with an associated increase in contact area. These results are reassuringly similar to previously observed helium-induced changes in tunsten[14] on which the hypothesis outlined above is based.

Cross-section TEM micrographs through the centre of a nano-indent in the implanted material clearly show defect free zones (Figure 1 (d)), inspiring confidence in the proposed mechanism of defect removal by glide dislocations in localised areas. These slip channels are found to be parallel to the [111] direction and are cleared of most visible defects. This observation is similar to slip channels observed in other irradiated materials [19,20,22,24,28,29]. In some cases, the experimentally observed slip channels appear clearly demarcated, with sharp boundaries separating the channels from the surrounding matrix [30]. Examples include neutron-irradated copper (irradiated at 300 K with $2.5 \times 10^{18}$ neutrons per cm$^2$) [19] and iron ion irradiated Fe-12Cr-20Mn (irradiated to a dose of $1 \times 10^{16}$ ions per cm$^2$). In contrast, the slip channels we observe have a less well-defined boundary with the surrounding, higher defect density material. A similar observation of slip bands with diffuse boundaries with the surrounding matrix was made in 0.012 dpa neutron-irradiated bcc vanadium [21]. In the case of vanadium, the diffuse boundary between the slip bands and the matrix, was caused by "weak channeling" i.e. a state close to the threshold dose for channeling, where there is sufficient defect density to initiate channeling, but where the channels are not completely defect free [21,27]. Accordingly, in vanadium exposed to higher damage levels of 0.12 and 0.69 dpa, the localised bands were found to become completely cleared of all defects, and more clearly demarcated from the surrounding matrix [21,27]. Similar observations of channels developing sharper, well defined boundaries with the surrounding matrix, with increasing dose, were seen in neutron irradiated pure iron (between 0.4 and 0.79 dpa) [28] and in neutron-irradiated zircaloy-4 (between 0.1 and 0.8 dpa) [22]. These previous observations imply that the 3000 appm helium-implanted tungsten studied here may be at the threshold dose for initating channeling, leading to the observation of the weak, less well defined channels in Figure 1 (d).

With regard to the channel morphology, we note that the channels in Figure 1 (d) are relatively straight, similar to those observed in irradiated fcc metals for all damage levels [19,20,24,29]. With increasing dose, one may expect that the straight slip channels in bcc tungsten will not only become wider, but also curved. For example in neutron irradiated bcc vanadium, wide, river-pattern shaped channels have been observed with increasing dose from 0.12 to 0.69 dpa [21,27]. Bending of channels was also observed in neutron irradiated bcc iron for increasing dose from 0.4 to 0.79 dpa [21]. The high stacking fault energy (SFE) in bcc metals is expected to promote cross-slip of screw dislocations [31], while in low SFE fcc metals, cross-slip is harder to initiate owing to the formation of partial dislocations. Multiple cross-slip events per dislocation can help the formation of curved channels in bcc metals as each cross-slip induces a sharp change in direction of the dislocation path [27]. Thus, the SFE may be considered responsible for differences in the slip channel morphology between bcc and fcc metals, particularly at high doses [27].

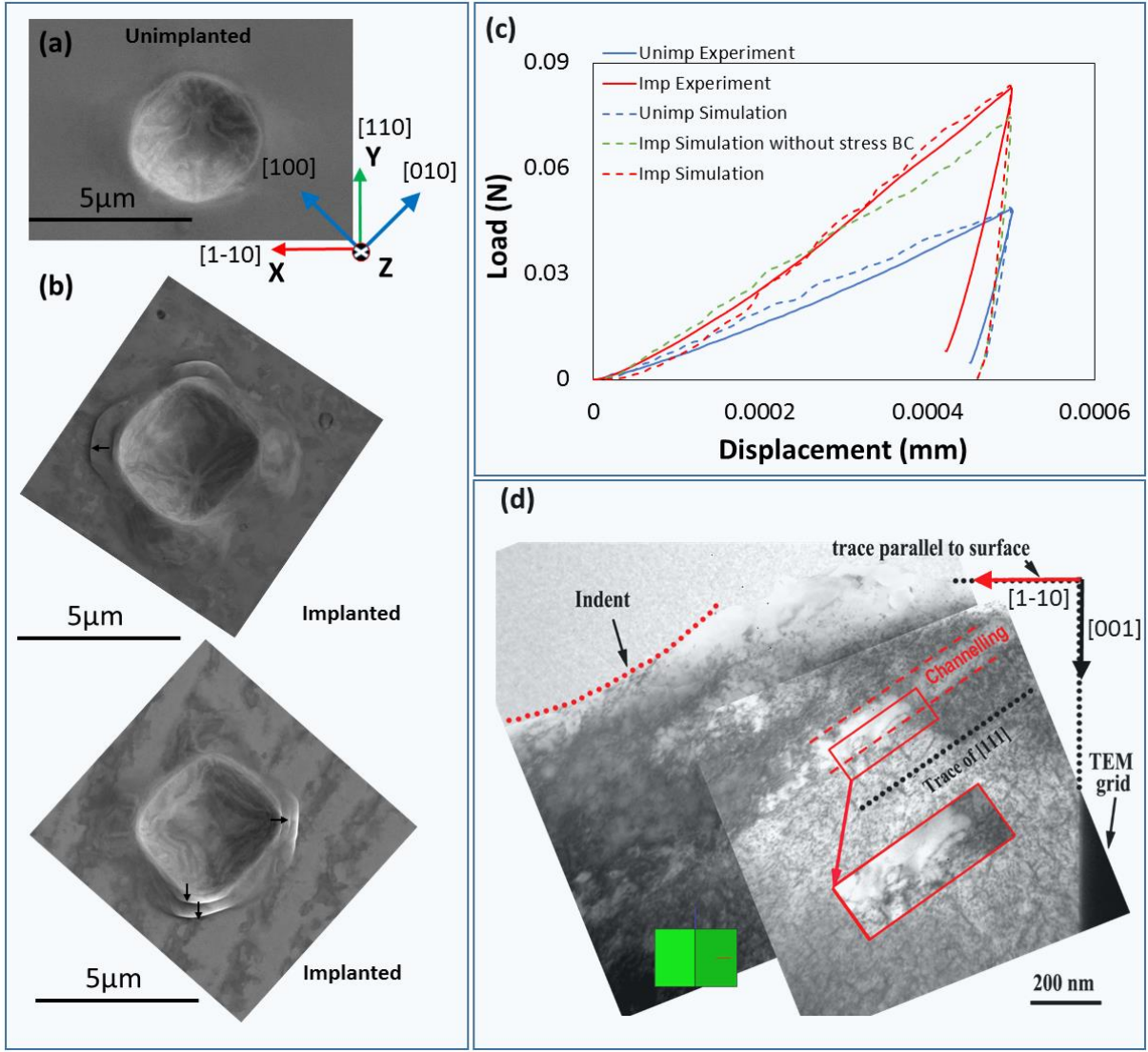

Figure 1 – (a) SEM micrographs of the nano-indent made with a 4.2 µm radius spherical indenter tip in the unimplanted part of sample and (b) SEM of two indents made in the helium-implanted part of the sample. The superimposed X, Y and Z axes apply to both (a) and (b). (c) Load versus displacement curve for the unimplanted and helium-implanted tungsten sample as obtained from nano-indentation and predicted by CPFE. (d) TEM micrographs of a FIB-lift-out cross-section from the indent-centre in the implanted sample, showing the presence of slip channels beneath the indent. The inset in (d) shows a close-up of the slip channel and depicts the diffuse boundary between the channel and the surrounding matrix. The green cube represents the crystal orientation of the TEM lamella, also indicated by red (X axis) and black (Z axis) arrows showing the same coordinate frame as in (a).

## 2.2. Developing the CPFE model

The proposed hypothesis for the interaction of glide dislocations with He$_2$V-SIA complexes was used to develop a user material (UMAT) for helium-implanted tungsten to allow CPFE simulations of nano-indentation of this material. Details of the UMAT code are provided in Section 4.6. Briefly, the UMAT uses a physically-based constitutive law governing the thermally activated glide of dislocations through a field of pinning dislocations and He$_2$V-SIA complexes. When the resolved shear stress on a given slip system $\tau^\lambda$ is greater than critically resolved shear stress $\tau_c$, slip is allowed. $\tau_c$ is modified incrementally as

$$\tau_c = \tau_c^0 + C'G\,b\,\sqrt{\rho_{GND}} + \tau_H \tag{1}$$

where the term $C'G\,b\,\sqrt{\rho_{GND}}$ accounts for strain hardening using a Taylor hardening law [32,33] and $\tau_H$ (with initial value $\tau_H^0$ at $t = 0$) accounts for the increased resistance to dislocation glide by the helium-implantation-induced defects. $G$ is the shear modulus of tungsten, $\tau_c^0$ the initial critically resolved shear stress (CRSS), $\rho_{GND}$ the sum of the GNDs produced on all slip-systems and $C'$ the proportion of GNDs considered contributing to hardening. For unimplanted pure tungsten, $\tau_c$ is evolved using Eq. (1), but without the $\tau_H$ term. Strain-softening in helium-implanted tungsten i.e. removal of He$_2$V-SIA complexes, is implemented by reducing $\tau_H$ incrementally

$$\tau_H^{t+\Delta t} = \tau_H^0 e^{-(\beta_p^{t+\Delta t}/\gamma)} \tag{2}$$

where $\beta_p^{t+\Delta t}$ is the accumulated crystallographic slip and $\gamma$ the softening rate. The rate of decrease of $\tau_H$ is estimated to be proportional to its current value i.e. the current helium-defect concentration $\left(\frac{\partial \tau_H}{\partial \beta_p} = -\tau_H/\gamma\right)$, suggesting the exponential softening rate in Eq. (2).

Only two UMAT parameters were fitted to the experimental result of the unimplanted sample; $\tau_c^0$ and $C'$. These two are essential for simulating nano-indentation of pure tungsten [34]. Of the two additional parameters introduced for helium-implanted tungsten, only $\gamma$ is fitted to the experimental result of the implanted sample. The fitted load-displacement curves from CPFE are superimposed as dotted lines on the experimentally obtained curves in Figure 1 (c)).

$\tau_H^0$ is computed as a function of the energy required for a gliding dislocation to dissociate helium from a He$_2$V-SIA complex and thus overcome this obstacle. $\tau_H^0$ as such depends on the concentration of such defect complexes, the size of an individual defect complex, the binding energy of the helium-vacancy complex and the migration energy of the released interstitial helium at room temperature (details in Section 4.6.3). The physically-based value of $\tau_H^0$ is a key novelty of the formulation, reducing the number of fitting parameters to only one ($\gamma$) for helium-implanted tungsten.

Additional boundary conditions, $\sigma_{xx}^{BC} = \sigma_{yy}^{BC} \approx$ -260 MPa are applied to the helium-implanted layer to account for the restricted in-plane displacement to maintain geometrical continuity between the implanted layer and substrate (calculations in Appendix A in Supplementary Material) [35]. It is interesting to compare the CPFE predicted load-displacement curves for the implanted material with and without these stress boundary conditions, keeping all other UMAT parameters the same. The indentation load predicted without the residual stresses is ~ 16% lower than the experimental observation. This implies that large residual stresses (here ~ 75% of $\tau_c^0$) must be accounted for in CPFE, to yield accurate predictions. Prior studies on hardness of mild steel, probed through micro-indentation, showed a similar dependence on residual stress state [36]. Interestingly our CPFE calculations highlight that the impact of residual stress on indentation response continues to be important even at the nano-scale.

### 2.3. Surface morphology of indents: AFM vs CPFE

The large increase in pile-up in the helium-implanted material predicted by CPFE agrees very well with the surface height profile measured by AFM (Figure 2). Line plots were made from AFM and CPFE contour plots in the <100> and the <110> directions i.e. direction with minimum and maximum pile-up respectively (Figure 2 (e)-(f)). From the line plots, ~ 172 % increase in pile-up height is seen in the <110> direction in the implanted sample. A relatively shallower residual indent impression in the implanted sample is consistent with the observed increase in hardness.

Observations of maximum and minimum pile-up in the <110> and <100> directions respectively, can be explained by considering the <111> slip directions in bcc tungsten. Dislocations gliding along any of the four <111> directions can exit where these directions intersect the sample surface (which here has [001] normal). As such, the pile-up on the [001] plane, is expected to occur preferentially along the projection of the <111> directions on the [001] plane i.e. along the four <110> directions lying in the [001] plane which are at 90° to each other (Figure 2). Little pile-up is expected to occur along the <100> directions that lie in the [001] plane. We note in passing that CPFE correctly predicts the experimentally observed four-fold pile-up pattern, by constraining slip to the applicable crystallographic slip systems. The contrasting, unphysical uniform distribution of pile-up around the indent, predicted by an isotropic continuum finite element (FEM) model, is displayed in Figure B.1 (Appendix B in Supplementary Material) for comparison.

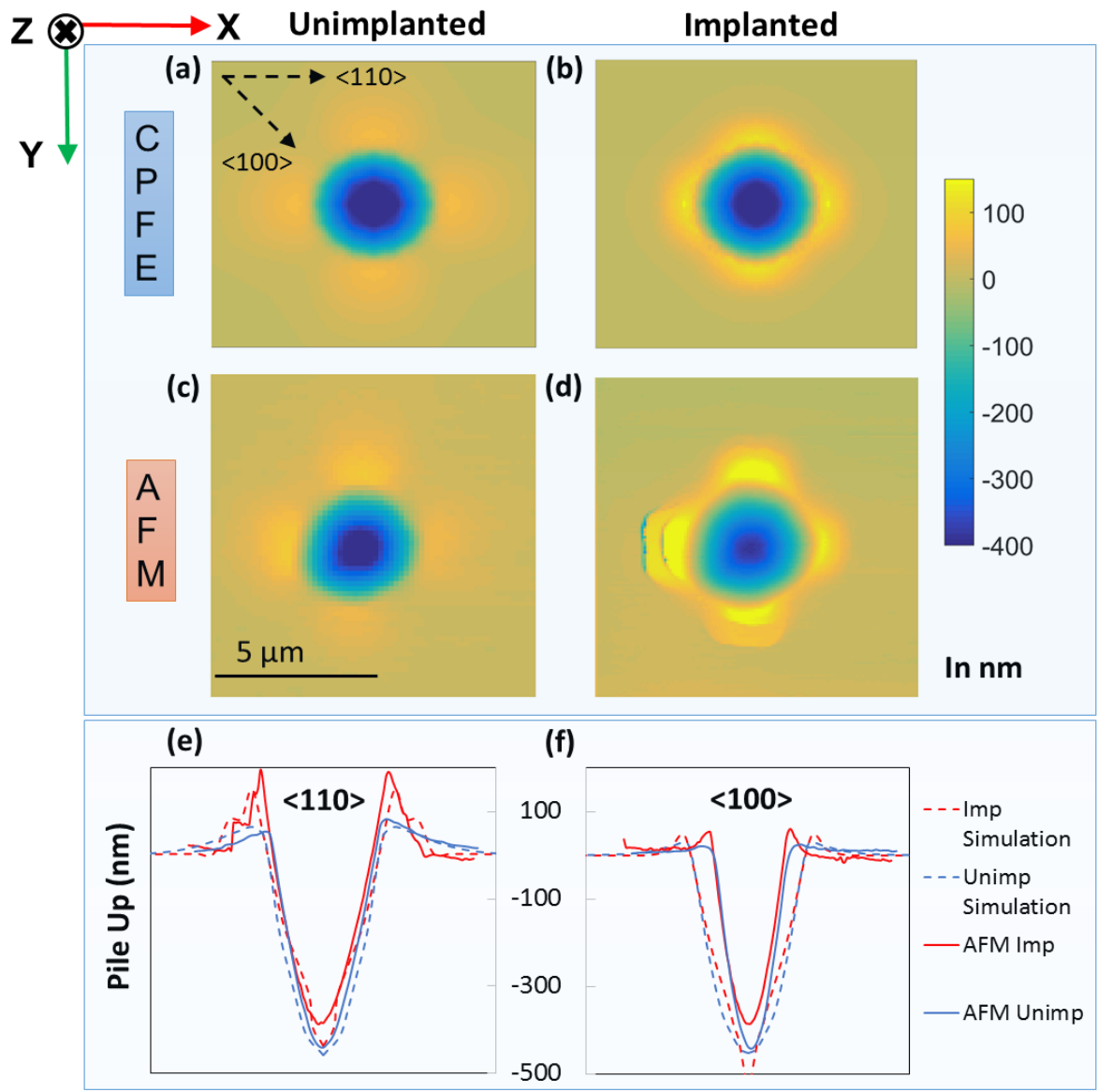

Figure 2 – Surface profiles of the nano-indents; (a) and (b) as obtained through CPFE formulation for the unimplanted and the helium-implanted tungsten sample respectively, (c) and (d) as obtained through AFM measurements for the unimplanted and helium-implanted tungsten sample respectively. (e) and (f) show line profiles of surface height extracted along <110> and <100> directions respectively for the unimplanted and implanted material. Also shown are the surface profiles predicted by CPFE.

### 2.4. Residual elastic lattice rotations

The residual elastic lattice rotations induced by nano-indentation around and beneath indents (represented by rotation matrix ***R***) were predicted by CPFE and compared with HR-EBSD and Laue diffraction measurements (Laue data used from prior study for comparison[14]).

The average orientation of points in the unimplanted material, far from the indent (22-25 μm and 10 μm below surface for Laue and HR-EBSD respectively) was considered as the orientation of the un-deformed material, $R_0$. For CPFE, the input orientation was considered as $R_0$. The indentation-induced change in crystal orientation from $R_0$ to $R$ at each point in space was then calculated as [34]:

$$R_t = R\,R_0^{-1} \qquad (3)$$

Using the convention in [37], $R_t$ accounts for the combined effect of sequential rotation about the X, Y and then Z axis by the rotation angles $\theta_x$, $\theta_y$ and $\theta_z$. Note, we use the common convention e.g. $\theta_x > 0$ if anticlockwise when viewed looking down the X-axis towards to origin (i.e. right handed rotations). The lattice rotations in Figure 3 are plotted on virtual YZ slices, at different X positions, for both samples. For HR-EBSD, only rotation about the out-of-plane axis $\theta_x$, is considered for one YZ slice through the indent centre. Rotations about the other two axes are not considered as they may have been modified due to stress-relaxation during FIB-lift-out preparation. HR-EBSD and Laue diffraction, respectively, have the advantage of high spatial resolution and high angular resolution, and are compared with the CPFE predictions using the same colour and length-scales.

The Laue data and CPFE predictions of lattice rotations (Figure 3 (a)-(d)) agree reasonably well, particularly for $\theta_x$. Interestingly, agreement is better for slices 5 μm away from the indent centre (positions 1 and 3 in Figure 3 (c)-(d)). The disparity between CPFE and Laue data in position 2 is due to the presence of steep strain gradients and rapid variation of lattice rotation at the indent centre. In positions 1 and 3, rotations in the unimplanted sample, particularly $\theta_x$, are significantly larger than in the implanted sample. For quantitative analysis, line plots were extracted, for these two slices, at depths of 4.5 μm from the sample surface, from both Laue and CPFE results and superimposed (Figure 4). The line plots distinctly show the smaller lattice rotations in the implanted sample, suggesting a more localised deformation.

The $\theta_x$ line plots from CPFE and Laue agree very well in the far field, where small changes in lattice orientation are well captured by the high angular resolution of Laue measurements [38–40]. However, in the highly deformed volume directly beneath the indent, where steep strain gradients exist, Laue peaks become streaked [41]. This streaking is linked to the underlying GND population within the probed volume, as described e.g. by Barabash et al. [42,43]. Unfortunately streaked Laue peaks cannot be fitted as accurately, leading to a loss of strain and lattice rotation resolution in this region. For example, in Figure 4, CPFE predicts rapid variations of lattice rotation immediately below the indent that are not resolved by Laue diffraction. The high spatial resolution of HR-EBSD [44,45] complements the Laue measurements by giving detailed information in this region.

Quite good agreement between HR-EBSD and CPFE (Figure 3 (e)-(f)) is seen for both samples in terms of the geometry of the deformation field. However, CPFE predicts a slightly wider spread of the rotation field than measured by HR-EBSD. This may be attributed to stress release during FIB-lift-out preparation. Importantly, comparison of CPFE and HR-EBSD shows that the magnitude of the rotation field immediately below the indent is similar for both the implanted and unimplanted material. Thus, the major difference between the implanted and unimplanted material lies in the far-field rotations captured by Laue diffraction. It is worth noting that using only HR-EBSD or Laue diffraction could lead to incomplete measurement of the deformation field in the far-field or near the indent respectively. This underlines the importance of the use of complementary techniques for gaining a complete picture of crystal scale deformation at the nano-scale.

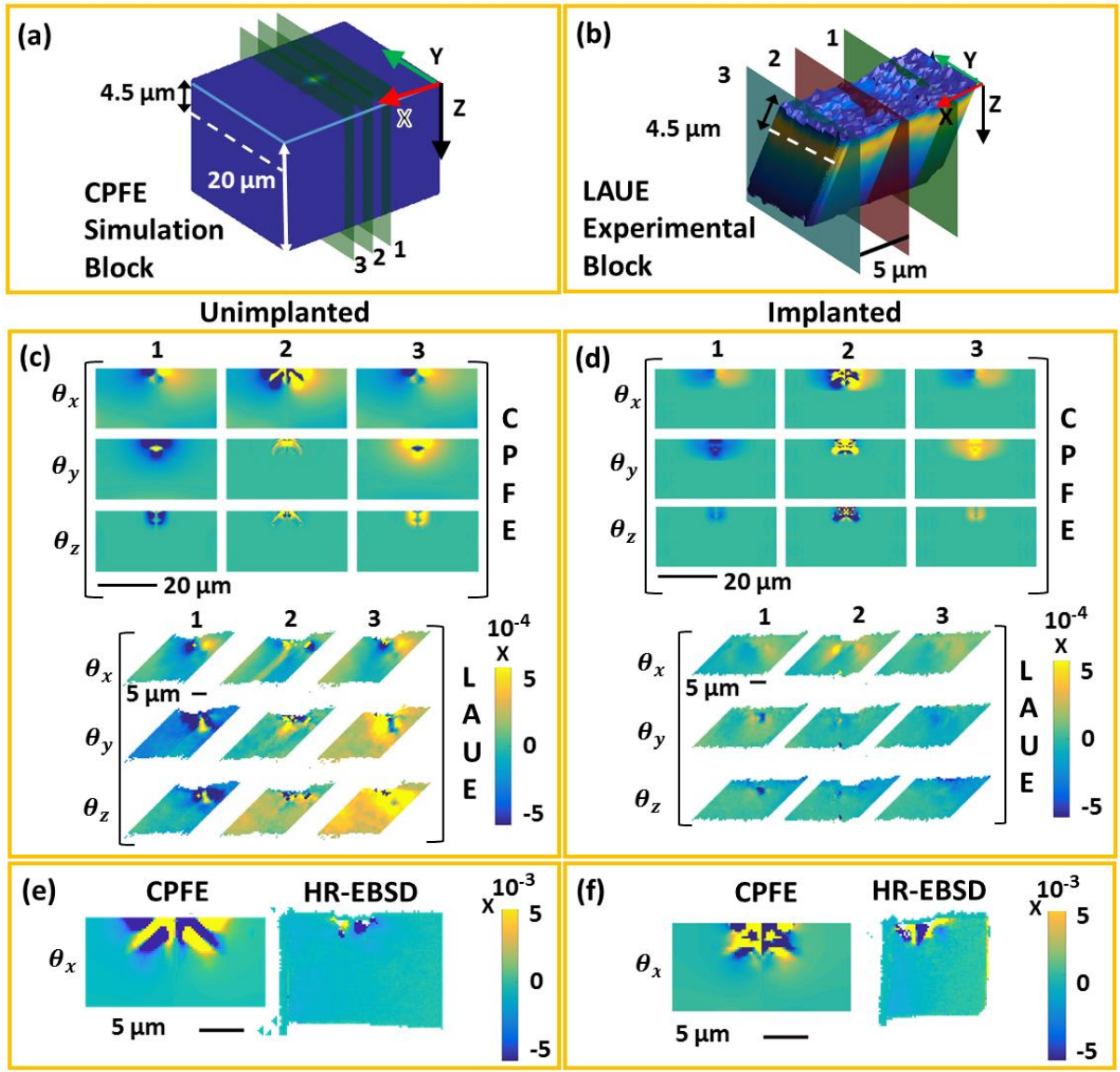

Figure 3 – (a) 3D rendering of the CPFE simulated volume coloured according to the predicted displacement magnitude along the Z axis of the indented tungsten block. X, Y and Z axes are superimposed. Slices 1-3, drawn on the block, represent the three sections along the X-axis (indent centre - slice 2), on which the lattice rotations predicted by CPFE are plotted. (b) Visualization of the sample volume measured by Laue diffraction, coloured according to the experimentally measured intensity. Superimposed are the X, Y, and Z axes, as well as the slices on which the measured lattice rotations are plotted. Comparison of lattice rotations predicted by CPFE and measured by Laue (Laue measurements in [14]) for the (c) unimplanted sample and (d) the helium-implanted sample. Comparison of lattice rotation as predicted by CPFE and measured by HR-EBSD for the (e) unimplanted sample and (f) the helium-implanted sample. With respect to the initial crystallographic coordinates, the X axis is [1 -1 0], Y is [1 1 0] and Z is [001]. Data from CPFE simulations, Laue diffraction and HR-EBSD experiments are displayed on the same length- and colour-scale. The dotted white lines in (a) and (c) represent the depth at which the line plots in Figure 4 were extracted. Lattice rotations are plotted in radians.

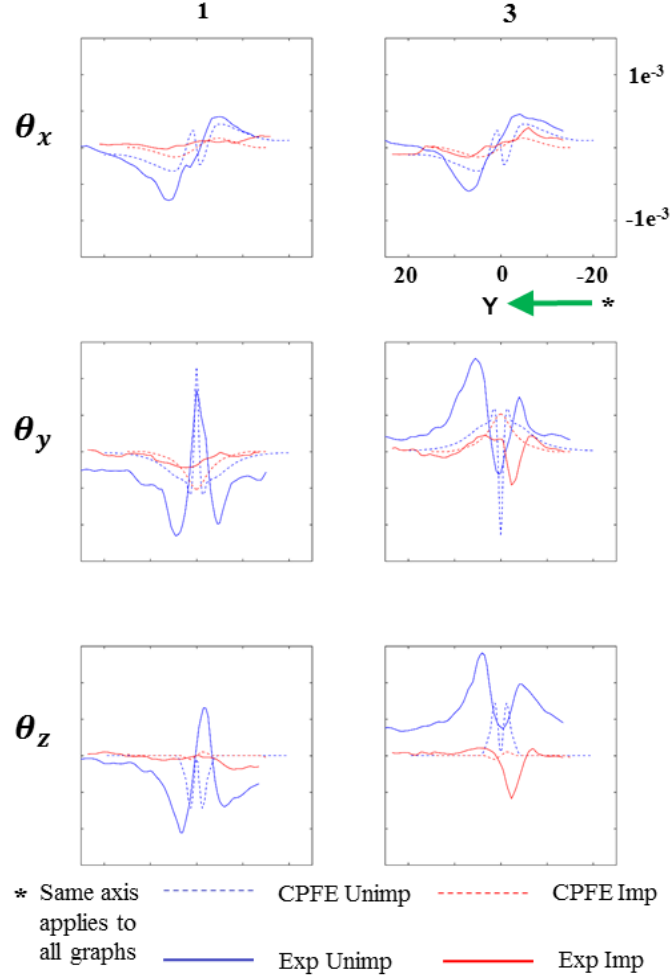

Figure 4 – Line plots corresponding to the contour plots in Figure 3 (c)-(d). Lattice rotations are plotted along a horizontal line 4.5 µm below the indent (line shown by dotted white lines in Figure 3 (a) and (b)). Slices 1 and 3 represent the YZ sections drawn 5 µm away from the indent centre, on either side, along the X-axis in Figure 3 (a) and (b).

### 2.5. Residual elastic lattice strains

The deviatoric residual lattice strain ($\boldsymbol{\varepsilon}^e_{dev}$) is measured by both Laue diffraction (Laue data used from prior study for comparison [14]) and HR-EBSD. To directly compare the CPFE predictions with the experimental measurements, $\boldsymbol{\varepsilon}^e_{dev}$ is calculated from the total elastic strain tensor ($\boldsymbol{\varepsilon}^e$) given by CPFE:

$$\boldsymbol{\varepsilon}^e_{dev} = \boldsymbol{\varepsilon}^e - \boldsymbol{\varepsilon}^e_{vol} = \boldsymbol{\varepsilon}^e - 1/3 \text{ Tr}(\boldsymbol{\varepsilon}^e)\boldsymbol{I} \qquad (4)$$

where $\boldsymbol{\varepsilon}^e_{vol}$ is the volumetric strain. Helium-implantation induces an out-of-plane lattice swelling i.e. positive $\varepsilon^e_{zz}$, while constraints imposed on the implanted layer along X and Y directions (to maintain continuity with the substrate) means that $\varepsilon_{xx} = \varepsilon_{yy} \approx 0$ [16,35]. As Tr $(\boldsymbol{\varepsilon}^e_{dev}) = 0$, it may be estimated that, far from the indent in the implanted material, $\varepsilon^e_{dev_{xx}} = \varepsilon^e_{dev_{yy}} = -0.5\varepsilon^e_{dev_{zz}}$. This implantation-induced deviatoric strain is clearly seen in Laue measurements of the implanted sample (Figure 5 (d)); a tensile $\varepsilon^e_{dev_{zz}}$ (~5.5 × 10$^{-4}$) and compressive $\varepsilon^e_{dev_{xx}}$ and $\varepsilon^e_{dev_{yy}}$ (~ -2.75 × 10$^{-4}$) in the implanted layer [14]. It is worth noting that the CPFE formulation for the implanted sample does not explicitly include the helium-induced lattice swelling. Rather the corrective stress, arising due to restricted deformation in the X and Y directions, is included in the CPFE model as an additional boundary condition (details in Appendix A in Supplementary Material). Consequently, the CPFE predicted strains are purely indentation-induced (direct components of the indentation-induced deviatoric strain referred to as $\varepsilon_{xx}$, $\varepsilon_{yy}$ and $\varepsilon_{zz}$). Thus, the CPFE predicted out-of-plane strain for the implanted layer is somewhat smaller than that measured by Laue, where both indentation and implantation induced strains are captured (Figure 5 (d)).

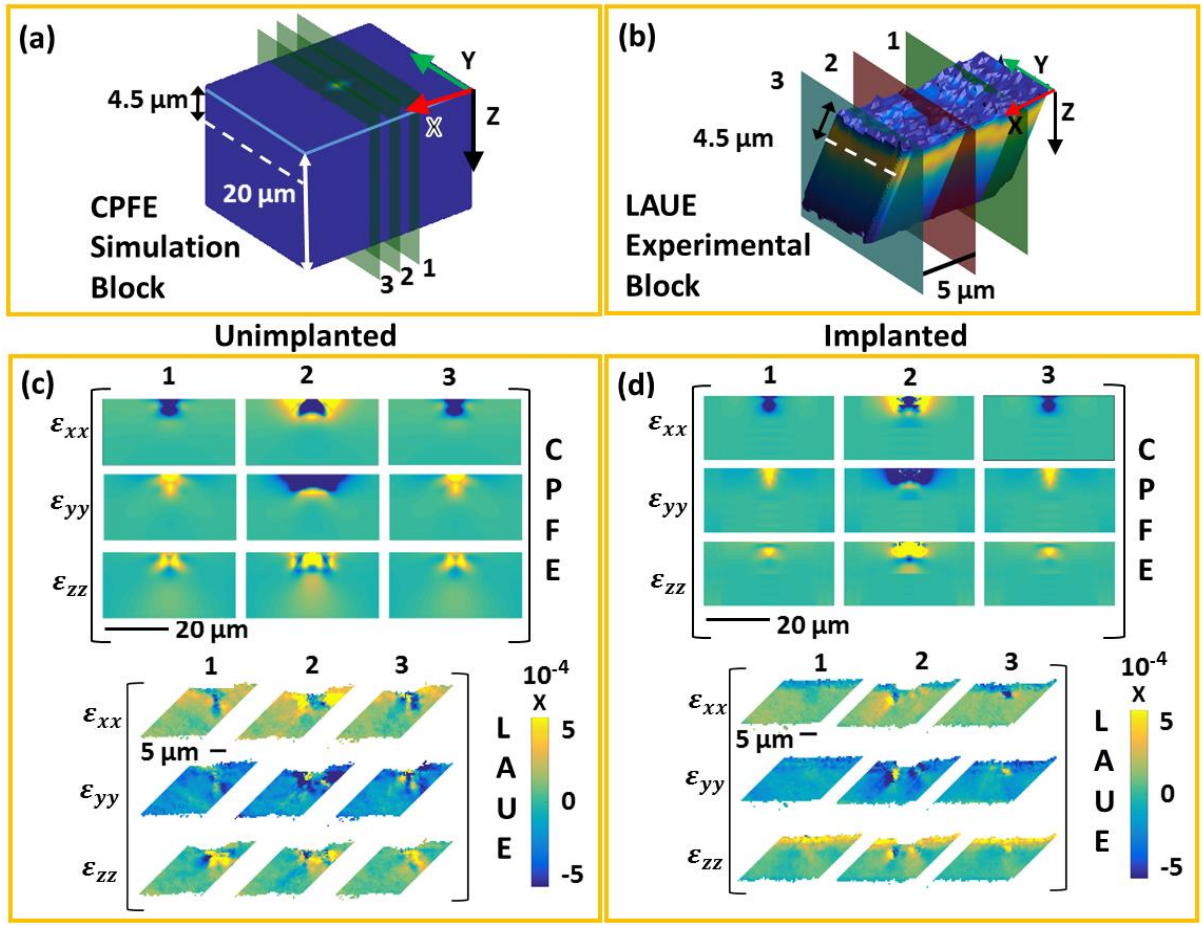

Figure 5 – **For the CPFE data**: (a) Slices 1-3 drawn on the block represent the three sections along the X-axis (slice 2 being at the indent centre) on which the residual deviatoric elastic strains predicted by CPFE are shown in (c) unimplanted sample and in (d) helium-implanted sample. **For the Laue data** (Laue measurements in [14]): (b) Visualisation of the measured sample volume. Superimposed are the X, Y, and Z axes, as well as the slices on which the measured deviatoric elastic lattice strains in (c) the unimplanted sample and (d) the helium-implanted sample are plotted. With respect to the initial crystallographic coordinates, X is [1 -1 0] direction, Y is [1 1 0] and Z is [001]. Slices 1-3 in experiments and simulations are at the same spatial positions and data in (c) and (d) is displayed on the same length- and colour-scales. The dotted white lines through (a) and (b) represent the depth at which the line plots in Figure C.1 in the Supplementary Material were extracted.

$\varepsilon_{xx}$, $\varepsilon_{yy}$ and $\varepsilon_{zz}$, predicted by CPFE and measured by Laue diffraction, are shown in Figure 5, plotted on virtual YZ slices through the sample at different X positions. Good qualitative agreement is observed. Interestingly both the measurements and CPFE predictions show a localisation of indentation induced lattice strain in the implanted sample. Distinctly noticeable is the tightly confined $\varepsilon_{zz}$, at positions 5 μm away from the indent centre (slices 1 and 3 in Figure 5 (d)). The localisation of strain fields is further quantified by line plots drawn for the corresponding contour plots in Figure 5 at depth of 4.5 μm from the sample surface (Appendix C in Supplementary Material). The agreement between CPFE predictions and experimental measurements inspires confidence in the CPFE formulation for helium-implanted tungsten (with only one fitting parameter) and its foundation hypothesis; localised deformation in slip channels arising from interactions of dislocations with $He_2V$-SIA complexes.

### 2.6. GND Density

Hardening is implemented in the CPFE formulation as a function of evolving GND density (Eq. (1)). For the implanted sample, the hardening is accompanied by a strain softening due to removal of helium-implantation-induced defects by glide dislocations (Eq. (2)). The fine-tuned combination of hardening and softening in the UMAT is vital for the quantitative agreement observed between the CPFE predictions and experimental results for the lattice distortions. The explicit incorporation of GNDs in the formulation is key for introducing length-scale effects as originally proposed by Ashby [46]. Further, the GND distribution can provide insights into the plastic zone size, which can be directly compared to experiments. Following Nye's theory [47], GND densities are thus computed from both HR-EBSD and Laue measurements and compared with the CPFE predictions.

Details of the GND density computation can be found elsewhere [34]. Here, the GND density $\boldsymbol{\rho}$ was computed using the L2 optimisation technique [48], which minimizes the sum of squares of dislocation densities i.e. $\sum_j \rho_j^2 = \boldsymbol{\rho}^T \cdot \boldsymbol{\rho}$. Previous work showed that L2, while mathematically simpler, can reliably predict the sum of GND densities over all slip systems [34]. We assume, consistent with the CPFE calculations, that deformation in bcc tungsten is accommodated by dislocations with a/2<111> Burgers' vector gliding on {110} planes [49,50] (list of Burgers' vector and line directions considered can be found elsewhere [34]). Furthermore, we assume dislocations to be of either pure edge (with <112> line directions) or pure screw type (with <111> line directions), resulting in 16 dislocation types in total [34].

Figure 6 shows the total GND density computed from HR-EBSD, Laue diffraction and CPFE datasets, plotted on the YZ, XZ and the XY cross sections at the indent centre. To enable a direct comparison with both Laue and HR-EBSD measurements, the CPFE predictions are plotted on two different length scales. Complementing each other, the high angular resolution of Laue measurements [38,51] and high spatial resolution of HR-EBSD measurements [44,52] detect the far-field and near-field deformation respectively. In the unimplanted sample, a small zone of high GND density is seen close to the indent (Figure 6 (d)), with a surrounding larger zone of low GND density (Figure 6 (c)). In contrast, the implanted sample shows a relatively larger zone of high GND density surrounding the indent (Figure 6 (h)), with a surprisingly sharp boundary and very low GND density outside this region (Figure 6 (g)). The experimental observations show a similar picture, consistent with the predictions from CPFE.

The CPFE formulation for the helium-implanted sample was based on the hypothesis of removal of helium-implantation-induced obstacles by the progressive passage of dislocations. This leads to defect-free slip channels that accommodate deformation in the implanted zone, near the indent, and give rise to a large pile up. As a result, a more compact plastic zone is expected in the implanted material, which is clearly seen in the spatial variation of GND density predicted by CPFE and observed in experiments. An important point to note from Figure 6 is the size of the plastic zone surrounding each indent. The results show that, although the indent is only 500 nm deep, the affected zone extends beyond the ~ 3 μm thick implanted layer. This implies that the underlying substrate can have a significant impact on the results, even when the indentation depth is < 20% of the implanted layer thickness. Generally, when nano-indentation is performed on thin film on substrates, the contribution of the underlying substrate is carefully accounted for [53–55]. However, in most studies investigating the indentation response of irradiated materials, indentation depth of ~ 20 % of the implanted layer thickness is used and the effect of the underlying unimplanted material is not considered explicitly [56–58]. Our result highlights that, to ensure fidelity, it is important to account for the combined effect of the implanted layer and substrate. An analysis whereby the implanted layer is approximated as an infinite half space may render inaccurate result.

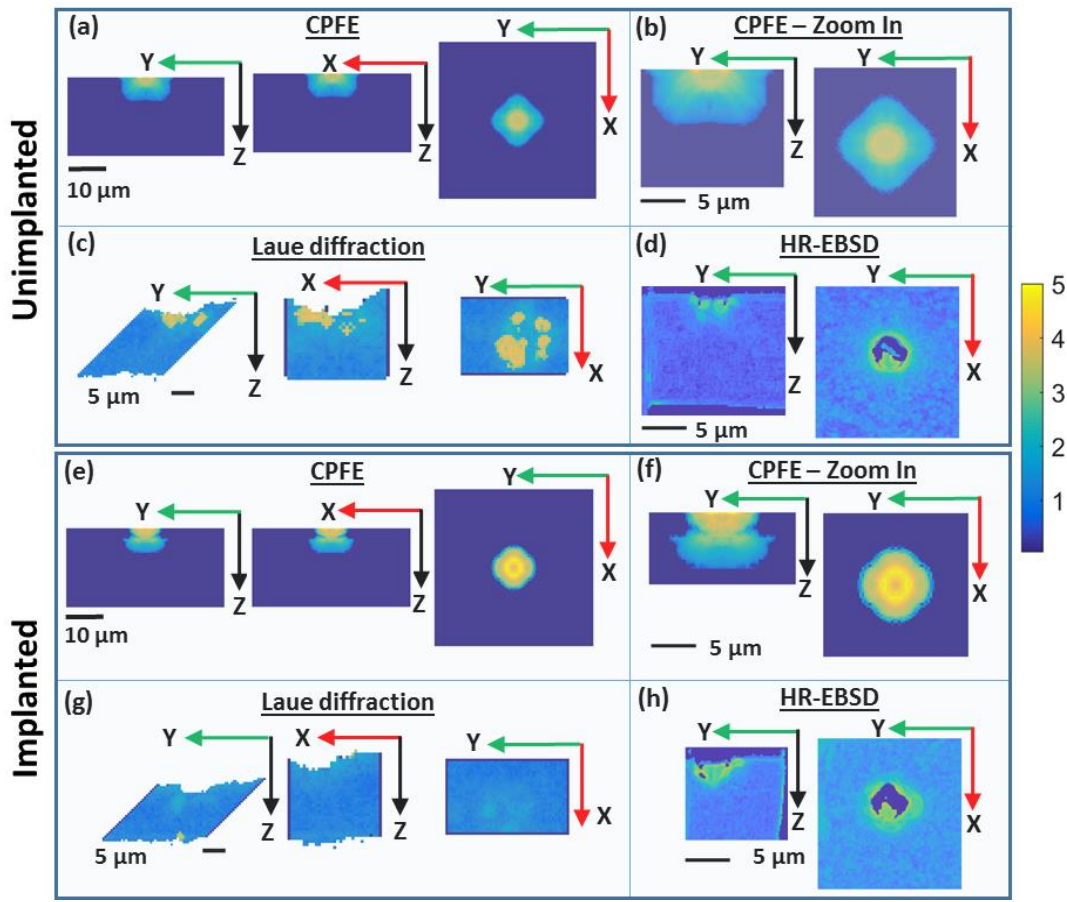

Figure 6 – Total GND density plotted on YZ, XZ and the XY cross-sections through the indent centre. GND density was calculated using L2 optimisation. Predictions from CPFE calculations are shown in (a) and (b) for the unimplanted sample, and (e) and (f) for the helium-implanted sample. GND density computed from Laue diffraction measurements for the unimplanted and implanted samples are shown in (c) [34] and (g) respectively. GND density determined by HR-EBSD for the unimplanted and the implanted sample is shown in (d) and (h) respectively. The same colour scale is used for all plots showing $\log_{10}(\rho)$ with $\rho$ in $1/\mu m^2$. (a), (c), (e) and (g) are plotted on the same length scale. (b), (d), (f) and (h) show magnified plots of GND density and are also all shown on the same length scale.

## 3. Conclusion

We have presented a comprehensive analysis of the effect of helium-implantation damage on the mechanical properties of tungsten. TEM of the cross-section of an indent in the implanted sample showed that dislocations "sweep out" helium-implantation-induced defects, leaving behind low defect density zones/channels. This "sweeping out" leads to a localisation of deformation and a softening of the material with increasing plastic deformation. Qualitatively this explains the substantial increase in pile-up and the localisation of deformation leading to slip steps. A physically-based CPFE model was constructed where initial hardening due to interaction of dislocation with irradiation defects, as well as subsequent softening is accounted for. Importantly this model has only one additional fitting parameter, for simulating helium-implanted tungsten, besides fitting parameters for CRSS and hardening co-efficient which are essential for simulating pure tungsten. It successfully captures the changes in load-displacement curve, changes in pile-up structure, lattice rotations and residual lattice strains, as well as GND population in remarkably good agreement with our experimental observations. The results highlight that even a comparatively small concentration of irradiation/implantation induces defects – 1500 appm – can produce dramatic changes in the mechanical behaviour, with strain hardening being replaced by a large increase in the initial hardness followed by strain softening. The accuracy of predictions of the CPFE model developed here inspires confidence that such physically-based models that capture the key mechanisms governing irradiation-induced changes, can be effectively used to determine the evolution of mechanical properties in fusion reactor armour components and assessing their in-service performance.

## 4. METHODS

### 4.1. Sample Preparation

A 2.5 mm radius, 0.8 mm thick disc with [001] out-of-plane orientation was cut from a tungsten single crystal (99.99% purity). A high-quality surface finish was obtained by mechanical grinding, polishing with diamond paste and final chemo-mechanical polishing with 0.1 µm colloidal silica suspension.

### 4.2. Helium Implantation

Implantation was performed at the National Ion Beam Centre, University of Surrey, UK. The sample was partly implanted with helium at 298 K, using a 2 MeV ion accelerator, and a raster scanned beam, to ensure a uniform implantation dose. Using a combination of different ion energies and fluences (Appendix D in Supplementary Material), a near uniform helium ion concentration of ~3000 appm within a ~2.8 µm thick implanted layer was obtained [16,59,60]. The SRIM estimated implantation profile [61] (displacement energy of 68 eV, single-layer calculation model [62]) is shown in Appendix D. A helium ion concentration of ~3100 ± 410 appm, with an associated damage of 0.20 ± 0.03 displacements per atom (dpa) was obtained between 0-2.8 µm depth. Frenkel pair formation is the main damage mechanism, since helium implantation induced recoils at energies <2 MeV have predominantly low energy [16]. Little defect clustering is expected, given that the mobility of vacancies at room temperature is low [11,63–65].

### 4.3. Nano-Indentation, AFM & SEM

Nano-indentation was performed on a MTS NanoXp, with a spherical diamond tip (~ 4.2 µm radius, 90° cone angle, from Synton MDP) to a maximum displacement of 500 nm with 50 µm spacing between indents. Indent impressions were imaged using a Zeiss Merlin FEG SEM. Surface morphology was determined using a Digital Instruments Dimension 3100 AFM in contact mode using Bruker CONTV-A tips (nominal tip radius 10 nm).

### 4.4. Micro-beam Laue diffraction

X-ray micro-diffraction was carried out at beamline 34-ID-E, Advanced Photon Source, Argonne National Lab, USA. A polychromatic X-ray beam (7-30 keV) was focused by KB mirrors to a sub-micron probe spot (500 nm (vertical) × 400 nm (horizontal) full width at half maximum near Lorentzian shape). With the sample placed at the probe focus in 45° reflection geometry, diffraction patterns were recorded on an area detector (Perkin-Elmer, #XRD 1621, with pixel size 200×200 µm) placed ~511 mm above the sample.

The recorded images correspond to the total intensity scattered by the entire volume illuminated by the incident beam. The depth along the incident beam from which a specific diffraction signal originated is thus unknown. At 34-ID-E, the Differential Aperture X-ray Microscopy (DAXM) technique enables depth-resolved measurements. Here, a platinum wire is scanned in small steps between the detector and the diffracting sample. The depth vs intensity profile for each detector pixel is recovered by subtracting the diffraction images from consecutive wire positions and triangulating using the wire edge and the line of the incident beam. Detailed description of the DAXM technique and the 34-ID-E instrument is given elsewhere [38,51,66]. For the present sample, material up to 20 µm beneath the ion-implanted surface was measured with 0.5 µm 3D spatial resolution. Laue diffraction patterns containing 30+ peaks were indexed and fitted using the LaueGo software package (J.Z. Tischler: tischler@anl.gov) to extract both lattice orientation and the full deviatoric lattice strain tensor at each measured point in 3D space.

### 4.5. FIB Lift-out, HR-EBSD & TEM

Focused ion beam (FIB) lift-out cross-section samples (~ 3 µm thick) were made from the indent-centre in the implanted and unimplanted samples using a Zeiss Auriga FIB-SEM. For HR-EBSD measurements, lift-outs were thinned from one side to ~1.5 µm thickness using a 30 kV, 240 pA ion beam, followed by low energy ion beam polishing (2 kV, 200 pA, 10 minutes) to remove FIB damage. For TEM characterization, the lift-out specimens were further thinned to ~100 nm using FIB, and then flash polished (1% NaOH aqueous solution, 0.05 s, 8 V, 298 K). Though the FIB lift-out preparation removes constraints perpendicular to the foil, the in-plane lattice strains and rotations are expected to remain largely unchanged.

HR-EBSD measurements were performed on a Zeiss Merlin FEG SEM equipped with a Bruker eFlash detector. The Kikuchi pattern at each point is recorded at the full detector resolution. The 2D spatial variation of lattice distortions is obtained by cross-correlating the EBSD pattern collected at each point in a map with a pattern collected at a nominally strain-free reference position (chosen far away from the indent (>10 μm) in the unimplanted substrate material) [67–69]. From this the deviatoric lattice strain (~1 × $10^{-3}$), lattice rotation (~1 × $10^{-3}$ rads) and the GND density for each point was found [47,48,70,71]. Analysis of the Kikuchi patterns was carried out using the XEBSD code provided by A. J. Wilkinson [44].

TEM characterization was performed on a JEOL 2100 TEM at 200 keV. The TEM micrographs were taken under a two-beam bright field condition.

### 4.6. CPFE Modelling

The indentation experiments were simulated using a strain-gradient CPFE model based on the model proposed by Dunne et al. [72,73] where plastic deformation is constrained to occur only in directions consistent with crystallographic slip. Recently we successfully demonstrated the use of strain-gradient CPFE for performing 3D simulations of nano-indentation in pure tungsten [34]. Here, we use a modified version of this UMAT to simulate nano-indentation in helium-implanted tungsten.

#### 4.6.1. The CPFE Model

A 3D model consisting of a 20×20×20 μm³ deformable block and a rigid 4.2 μm radius spherical indenter was simulated using Abaqus 2016 (Dassault Systèmes, Providence, RI,

USA) (Figure E.1, Appendix E in Supplementary Material). A refined finite element mesh (applied edge bias 0.1 to 2 µm) with 39500, 20-noded, reduced integration (8 integration points) 3D quadratic elements (C3D20R) was used. The 20 µm high simulation block was partitioned into two layers: a 3 µm thick implanted layer at the free surface and a 17 µm thick substrate. Considering the four-fold symmetry of <001>-oriented grains [70], only one quarter of the experiment was simulated. Mirror boundary conditions were applied on the XZ and YZ planes. The top surface was traction free and remaining surfaces of the sample block were fixed. Additional boundary conditions, $\sigma_{xx}^{BC} = \sigma_{yy}^{BC} \approx$ -260 MPa are applied to the helium-implanted layer to account for the restricted in-plane displacement to maintain geometrical continuity between the implanted layer and substrate (calculations in Appendix A in Supplementary Material) [35]. The indenter was subjected to a displacement of 0.5 µm into the sample block before unloading. Further details about the model are given in Appendix E in Supplementary Material.

### 4.6.2. The CPFE Framework

Crystal plasticity was implemented using a user material subroutine (UMAT) that shares data between gauss points using a common block. The UMAT code is based on the user element (UEL) originally developed by Dunne et al. [73]. Detailed description of this can be found elsewhere [34]. Briefly, a physically based slip law [73] was used to approximate the crystallographic slip rate $\dot{\beta}_p^\lambda$ for each slip system λ, by considering the thermally activated glide of mobile dislocations $\rho_m$ in a field of pinning dislocations. If the magnitude of $\tau^\lambda$, the resolved shear stress on slip system λ is less than $\tau_c$, the critically resolved shear stress, then $\dot{\beta}_p^\lambda$ is set to zero, as there is no change in plastic deformation. When $\tau^\lambda$ is found to be greater than $\tau_c$ the slip rate is estimated as

$$\dot{\beta}_p^\lambda(\tau^\lambda) = \rho_m \nu (b^\lambda)^2 \exp\left(-\frac{\Delta F}{kT}\right) \sinh\left(\frac{sgn(\tau^\lambda)(|\tau^\lambda| - \tau_c)V}{kT}\right) \quad (5)$$

where $\nu$ is attempt frequency, $b^\lambda$ the Burgers vector magnitude, $\Delta F$ the activation energy, $k$ the Boltzmann constant, $T$ the temperature and $V$ the activation volume, which depends on the spacing between the pinning dislocations $l$. We assume $l = \frac{1}{\sqrt{\Psi \rho_{SSD}}}$, where, coefficient $\Psi$ represents probability of pinning and $\rho_{SSD}$, the density of statistically stored dislocations (SSD).

Values of the UMAT parameters are listed in Appendix F of the Supplementary Material. In the simulation we make assumptions of isotropic elasticity, small elastic deformations and that all parameters on the RHS in Eq. (5) remain constant (other than $\tau_c$ and the independent variable $\tau^\lambda$). Evolution of $\tau_c$ in the unimplanted and the helium-implanted layer and the strain-softening for the implanted layer is done as described in Section 2.2.

### 4.6.3. Fitting Parameters for the UMAT

Only two UMAT parameters, $\tau_c^0$, and $C'$, were fitted to the experimental results of the unimplanted sample. When simulating indentation of the unimplanted material, both layers of the block were assigned identical material properties including the two fitting parameters $C'$ and $\tau_c^0$. To simulate helium-implanted tungsten, the 17 µm thick substrate was left unaltered and the implanted layer was assigned material parameters of the substrate along with the two additional parameters, $\gamma$ and $\tau_H^0$. Only $\gamma$, is fitted to the experimental result of the implanted sample while $\tau_H^0$ is physically-based.

To determine $\tau_H^0$ a slip-plane interspersed with helium-implantation induced defects is considered (Appendix G in the Supplementary Material). In Figure G.1, each circle represents a He$_2$V-SIA complex, separated from each other by a spacing $L$ and resisting dislocation glide with force $F_s$. Thus, in order to glide, the force per unit length required on the dislocation must be

$$\tau_H^0 b = \frac{F_S}{L} \quad (6)$$

An area $L^2$ of the slip plane contains one defect. This area also contains $(L/b)^2$ atoms in total, so knowing defect concentration $c$ Eq. (6) can be re-written as

$$\tau_H^0 = \frac{F_S}{bL} = \frac{F_S\sqrt{c}}{b^2} \quad (7)$$

The force $F_S$ may be considered a function of the energy required for the dissociation of helium from He$_2$V-SIA complex i.e.

$$2r_W F_S \approx \Delta E \quad (8)$$

where, $\Delta E$ is the dissociation energy (which includes the binding energy for the helium-vacancy complex and the migration energy of the released helium ion) and $r_W$ the atomic radius. The defect complex, primarily involving nearest neighbours [18], is estimated to be twice the size of the atomic radius of tungsten. Based on ab-initio studies of the stability of He$_2$V, $\Delta E$ is estimated to be approximately 2.5 eV [74]. Given the helium-implantation dose of 3000 appm, the concentration of defects, $c$, is estimated as $1.5 \times 10^{-3}$ (assuming on average 2 helium atoms per Frenkel pair) [16]. Knowing the values of $b$, $\Delta E$, $r_W$ (135 pm [75]) and $c$, $\tau_H^0$ can be estimated as

$$\tau_H^0 = \frac{\Delta E \sqrt{c}}{2r_W b^2} = 750 \text{ MPa} \quad (9)$$

## Acknowledgements

We thank N. Peng for performing the ion-implantation, M. Rieth for providing the tungsten single crystal sample, R. Xu and W. Liu for their help with X-ray diffraction experiments, A. London and D. Nguyen-Manh for their insightful comments, and A.J. Wilkinson for providing the software for HR-EBSD analysis. This work was funded by Leverhulme Trust Research Project Grant RPG-2016-190, and used resources of the Advanced Photon Source, a U.S. Department of Energy (DOE) Office of Science User Facility operated for the DOE Office of Science by Argonne National Laboratory under Contract No. DE-AC02-06CH11357. Ion implantations were performed under the UK Engineering and Physical Sciences Research Council grant EP/H018921/1. ET acknowledges financial support from the Engineering and Physical Sciences Research Council (EPSRC) Fellowship grant EP/N007239/1. Electron and atomic force microscopy were performed at the David Cockayne Centre for Electron Microscopy, Department of Materials, and at the LIMA lab, Department of Engineering Science, both at the University of Oxford.

**DATA AVAILABILITY**

The datasets generated during and/or analysed during the current study are available to download from https://doi.org/10.5281/zenodo.3237139.

**AUTHOR CONTRIBUTIONS**

SD conducted the nano-indentation, SEM and AFM experiments, did formal analysis of related data and the Laue diffraction data, developed and simulated the CPFE formulations, and wrote the original manuscript. HY conducted the SEM experiments, made the FIB lift-out sample, performed TEM and HR-EBSD measurements and edited and reviewed the manuscript. ET supervised development of the CPFE formulation, edited and reviewed the manuscript. FH performed the Laue diffraction experiments, supervised the overall project, edited and reviewed the manuscript. ET and FH both helped in funding acquisition.

# COMPETING INTERESTS STATEMENT

The authors declare no competing interest.

# Supplementary Material: Hardening and Strain Localisation in Helium-Ion-Implanted Tungsten

Suchandrima Das[a,1], Hongbing Yu[a], Edmund Tarleton[a,b,2], Felix Hofmann[a,3]

[a]Department of Engineering Science, University of Oxford, Parks Road, Oxford OX1 3PJ, UK
[b]Department of Materials, University of Oxford, Parks Road, Oxford OX1 3PH, UK

[1]suchandrima.das@eng.ox.ac.uk

[2]edmund.tarleton@eng.ox.ac.uk

[3]felix.hofmann@eng.ox.ac.uk

## Appendix A - Determination of the value of $\sigma_{xx}^{BC}$ and $\sigma_{yy}^{BC}$

The $\varepsilon_{zz}^{dev}$ component of the deviatoric lattice strain in the helium-implanted sample, as measured by the white-beam Laue diffraction, is ~ $550 \times 10^{-6}$. This value was obtained by averaging the measured $\varepsilon_{zz}^{dev}$ over 1.5 µm of the helium-implanted layer. The $\varepsilon_{zz}$ component of the total strain tensor was then computed from $\varepsilon_{zz}^{dev}$. The total strain tensor, $\boldsymbol{\varepsilon}^e$, is related to the deviatoric component $\boldsymbol{\varepsilon}_{dev}^e$ as

$$\boldsymbol{\varepsilon}^e = \boldsymbol{\varepsilon}_{dev}^e + 1/3 \; \text{Tr}\,(\boldsymbol{\varepsilon}^e)\boldsymbol{I} \qquad (A.1)$$

where $\boldsymbol{I}$ is the identity matrix. The $\varepsilon_{xx}$ and $\varepsilon_{yy}$ components of the total lattice strain tensor are expected to be zero, as deformation along X and Y directions is restricted in order to maintain geometrical continuity between the implanted layer and the substrate [1]. Thus $\text{Tr}\,(\boldsymbol{\varepsilon}^e) = \varepsilon_{zz}$. Therefore, Eq. (A.1) can be re-written as

$$\varepsilon_{zz} = \varepsilon_{zz}^{dev} + 1/3 \; \varepsilon_{zz} \qquad (A.2)$$

$$\varepsilon_{zz}^{dev} = 2/3 \; \varepsilon_{zz} \qquad (A.3)$$

Using Eq. (A.3), $\varepsilon_{zz} = 825 \times 10^{-6}$.

$\varepsilon_{zz}$ can be related to the "correctional boundary conditions" $\sigma_{corr_{xx}}$ and $\sigma_{corr_{yy}}$, generated to

counteract the helium-implantation induced eigenstrain, in order to maintain geometrical continuity between the implanted layer and the substrate [1].

$$\sigma_{corr_{xx}} = \frac{-E\varepsilon_{zz}}{3(1+\nu)} = \sigma_{corr_{yy}} \qquad (A.4)$$

Detailed derivation of Eq. (A.4) can be found elsewhere [1]. Knowing, $\varepsilon_{zz}$, the stress boundary conditions, $\sigma_{corr_{xx}} = \sigma_{corr_{yy}}$ or what is here referred to as $\sigma_{xx}^{BC} = \sigma_{yy}^{BC}$ are computed to be -260 MPa.

## Appendix B: Comparison of CPFE with Continuum-mode FEM

The CPFE predicted surface pile-up around the nano-indent in the helium-implanted material agrees very well with the AFM measurements quantitatively and in terms of pile-up patters (Figure 2 in main text). CPFE uniquely allows this prediction of the deformation pattern by accounting for crystallographic slip. A continuum plasticity simulation, performed on the same model, is unable to reproduce the orientation dependence of pile-up as shown in Figure B.1.

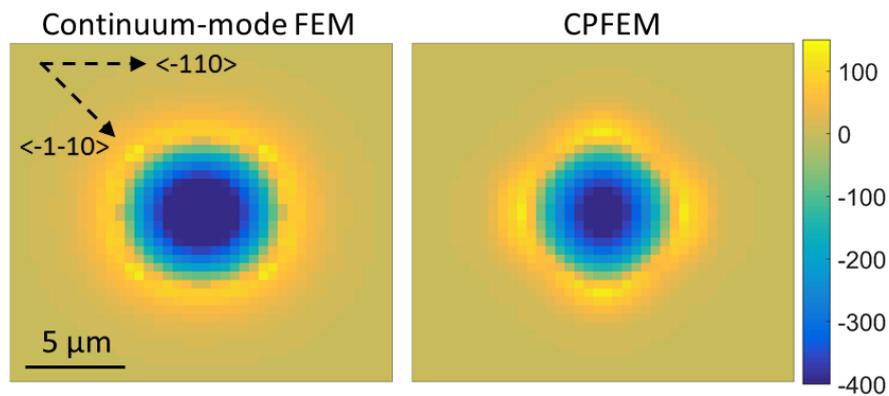

Figure B.1 – Surface pile-up around nano-indents in 001-oriented helium-implanted tungsten single crystal as predicted by continuum-based simulation (left, which does not account for crystal orientation) and by CPFE (right). The colour bar shows surface height in nm.

Appendix C

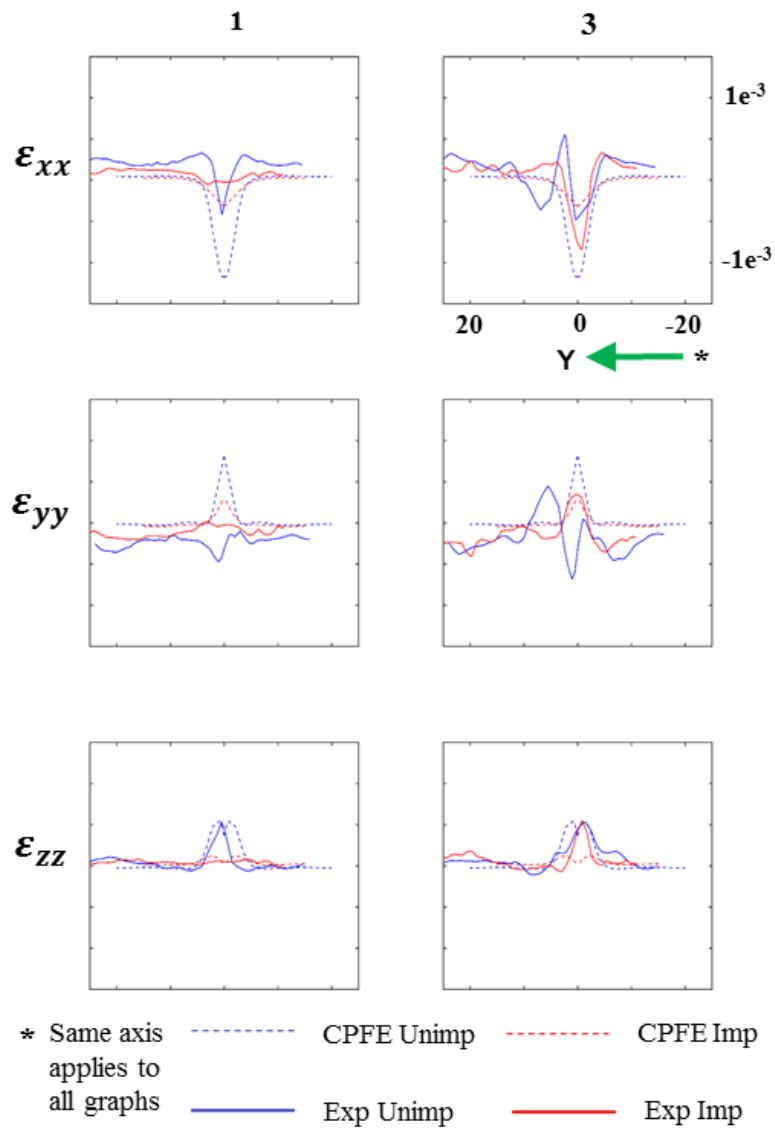

Figure C.1 - Line plots corresponding to the contour plots in Figure 5 (c) and (d) in the main text. Direct lattice strains are plotted along a horizontal line 4.5 µm below the indent (shown by dotted white lines in Figure 5 (a) and (b) in main text). Slices 1 and 3 correspond to YZ sections drawn 5 µm away from the indent centre, on either side, along the X-axis.

## Appendix D

List of 13 ion energies used and the corresponding fluence for the helium ion implantations:

| Ion Energy (MeV) | 3000 appm He Fluence (ions/cm²) |
|---|---|
| 0.05 | 2.40E+15 |
| 0.1 | 1.80E+15 |
| 0.2 | 4.20E+15 |
| 0.3 | 1.20E+15 |
| 0.4 | 4.80E+15 |
| 0.6 | 5.20E+15 |
| 0.8 | 5.00E+15 |
| 1 | 5.00E+15 |
| 1.2 | 5.00E+15 |
| 1.4 | 5.00E+15 |
| 1.6 | 5.50E+15 |
| 1.8 | 7.00E+15 |
| 2.0 | 5.00E+15 |

Table D.1 - List of 13 ion energies used and the corresponding fluences for the helium ion implantations.

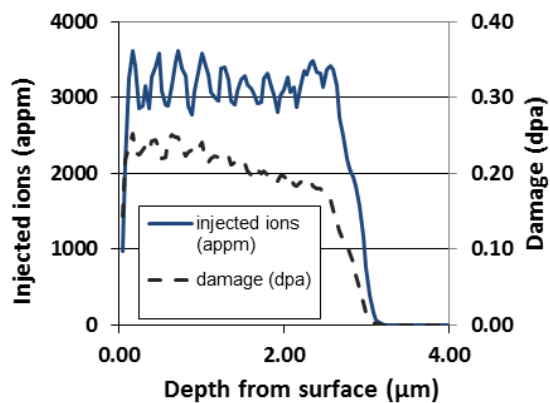

Figure D.1 - Helium-implantation profile as estimated by the SRIM code [2].

## Appendix E – Details of the CPFE Model

The sample block was assigned elastic properties of tungsten and the indenter was approximated as a discrete rigid wire frame (to avoid a full meshing and increase in simulation size). Contact between the sample and indenter was defined using the Abaqus node to surface contact algorithm and the contact was considered to be frictionless as past studies have shown that the mechanical response of the underlying substrate is not significantly affected by the coefficient of friction [3]. The results of indentation load reached were scaled with an effective modulus $E_{eff}$ to account for the indenter tip compliance [4].

$$P\ (exp.) = \frac{E_{eff\ (exp.)}}{E\ (FEA)}\ P(FEA) \qquad (E.1)$$

| $E_{diamond}$ | $E_{tungsten}$ | $\nu_{diamond}$ | $\nu_{tungsten}$ | $E_{eff}$ | $R_{indenter}$ |
|---|---|---|---|---|---|
| 1143 GPa | 410 GPa | 0.0691 | 0.28 | 322.58 GPa | 4.2 µm |

Table E.1 – Values of Young's modulus and Poisson's ratio for diamond (indenter tip) and tungsten (indented sample) as obtained from literature [5–7]. With the assumption of isotropic, linear elastic solid, the Young's modulus and Poisson's ratio are related to the elastic constant as follows: $E = c_{11} - 2\left(\frac{c_{12}^2}{c_{11}+c_{12}}\right)$ and $\nu = c_{12}/(c_{11}+c_{12})$.

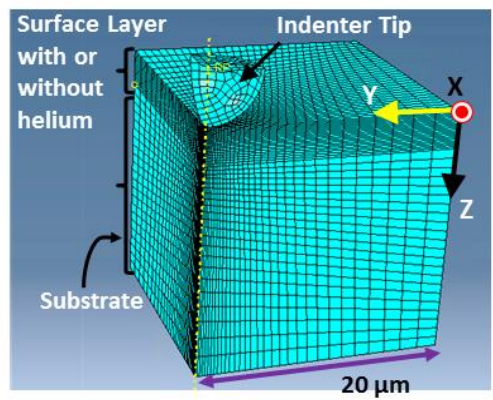

Figure E.1 - Refined mesh for the 3D crystal plasticity finite element simulation of the tungsten sample indented by a 4.2 µm radius spherical indenter with the X, Y, Z coordinate frame superimposed.

## Appendix F – UMAT Parameters

| Material Property | Value | Reference |
|---|---|---|
| Elastic modulus $E$ | 410 GPa | 5–8 |
| Shear modulus $G$ | 164.4 GPa | 5–8 |
| Poisson's ratio $v$ | 0.28 | 5–8 |
| Burgers' vector $b$ | $2.7 \times 10^{-10}$ m | 9 |
| Stress boundary conditions $\sigma_{xx}^{BC} = \sigma_{yy}^{BC}$ | -260 MPa | Appendix A |
| $\tau_H^0$ | 750 MPa | Section 4.6.3 in main text & Appendix G |
| Activation energy $\Delta F$ | 0.22 eV | Chosen to reduce strain-rate sensitivity |
| Boltzmann constant $k$ | $1.381 \times 10^{-23}$ J/K | 10 |
| Temperature $T$ | 298 K | Room temperature assumed similar to experimental conditions |
| Attempt frequency $v$ | $1 \times 10^{11}$ s$^{-1}$ | 11 |
| Density of statistically stored dislocations, $\rho_{SSD}$ | $1 \times 10^{10}$ m$^{-2}$ | Appendix H |
| Density of mobile dislocations $\rho_m$ | $3.5 \times 10^{10}$ m$^{-2}$ | Appendix H |
| Probability of pinning $\Psi$ | $0.657 \times 10^{-2}$ | Value chosen and kept fixed |
| $\tau_c^0$ | 360 MPa | Fitted to experimental data of unimplanted sample |
| $\gamma$ | 0.025 | Fitted to experimental data of helium-implanted sample |
| $C'$ | 0.0065 | Fitted to experimental data of unimplanted sample |

Table F.1– List of parameters used in the constitutive law in the CPFE formulation and their corresponding values.

## Appendix G: Determination of the value of $\tau_H^0$

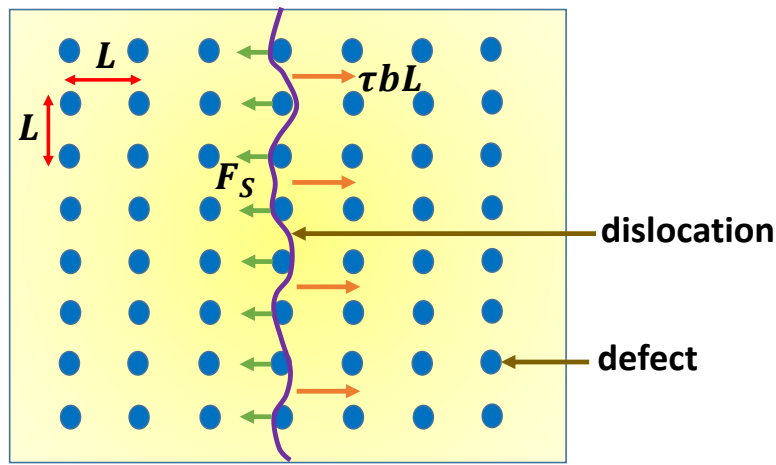

Figure G.1 – View of the slip plane interspersed with helium-implantation induced defects with a dislocation line trying to pass through.

## Appendix H: Estimating values for SSD

Laue diffraction measurements performed on nano-indented pure tungsten showed the presence of GND density on the order of ~ $10^{18}$ m$^{-2}$. This value agreed well with CPFE predictions of the indentation experiment [17]. The SSD densities assumed here, $\rho_{SSD}$ and $\rho_m$, for the CPFE calculations were taken to be much smaller than the GND density estimate, on the order of $10^{10}$ m$^{-2}$.